\documentclass{elsart}
\usepackage{amssymb}
\usepackage{natbib}
\usepackage{amsmath}

\input{tcilatex}

\begin{document}

\begin{frontmatter}

\title{Free Entanglement Measure of Multiparticle Quantum States}
\author{Chang-shui Yu \corauthref{yu}}
\author{, He-shan Song}
\corauth[yu]{Corresponding author.quaninformation@sina.com}
\address{Department of Physics, Dalian University of Technology,
Dalian 116024, China}

\begin{abstract}
In this paper, based on the classfication of multiparticle states and the
original definition of semiseparability , we give out the redefinition of
semiseparability and inseparability of multiparticle states. By virtue of
the redefinition, entanglement measure of multiparticle states can be
converted into bipartite entanglement measure in arbitrary dimension in
mathematical method. A simple expression of entanglement measure is given
out. As examples, a general three-particle pure state and an N-particle
mixed state are considered.
\end{abstract}

\end{frontmatter}

\section{\protect\bigskip Introduction}

Entanglement as a valuable resource has been widely applied to quantum
communication and quantum information processing. Quantum teleportation [1],
entanglement swapping [2], quantum key distribution [3] and quantum
correction and so on make use of quantum entanglement, the profoundly
important resource, in essence. Therefore, the quantification of
entanglement as a central problem in quantum information theory is a primary
goal of this field.

Quantum entanglement has attracted a lot of attention in recent years. A lot
of studies of quantum entanglement has been proposed and at the same time,
many entanglement measures come up [4-9]. However, only bipartite
entanglement with two levels has been perfectly complished [5], and there
are a lot of open questions in quantifying entanglement for the bipartite
entanglement measure in arbitrary dimensions and the multiparticle
entanglement measure. Fortunately, the method for classifying a
three-particle state [10] and the one presented recently for quantifying the
bipartite entanglement in arbitrary dimension [11] would increase our
understanding of multiparticle entanglement.

For some entanglement measure, there must exist a corresponding separability
criterion. However, so far there have not been an operational multiparticle
full separability criterion, but only a semiseparability definition [12].
Hence it is very difficult to obtain a thorough entanglement measure for
multiparticle systems.

In this paper, we study multiparticle free entanglement measure [12] with a
new idea(Strictly, free entanglement here denotes the entanglement of states
which excludes the fourth class defined in [10], and for convenience, free
entanglement is substituted by entanglement for multiparticle systems
later). Based on the classfication of multiparticle states, we express
semiseparability condition and full inseparability condition of
multiparticle systems in a unified way. By virtue of the conditions, we find
some mathematical counterpart of entanglement and convert multiparticle
entanglement measure into bipartite entanglement in arbitrary dimension in
mathematics. Then we give out a simple multipartitle entanglement measure
according to the bipartite quantum entanglement measure. Finally, we give an
example demonstrating our measure can work effectively in its right for pure
states and mixed states respectively.

\section{Separability and Inseparability}

We begin with the usual definition of multiparticle entanglement. For $N$%
-particle pure state $\left| \Psi ^{ABC\cdots N}\right\rangle $, if it can
be written in the form of direct product of all the subsystems, i.e.%
\begin{equation}
\left| \Psi ^{ABC\cdots N}\right\rangle =\left| \psi ^{A}\right\rangle
\otimes \left| \psi ^{B}\right\rangle \otimes \cdots \otimes \left| \psi
^{N}\right\rangle ,
\end{equation}%
then $N$-particle pure state is separable; If $N$-particle mixed state $\rho
^{ABC\cdots N}$ is separable, the state can be written in the following form:%
\begin{equation}
\rho ^{ABC\cdots N}=\underset{i}{\sum }p_{i}\left| \psi
_{i}^{A}\right\rangle \left\langle \psi _{i}^{A}\right| \otimes \cdots
\otimes \left| \psi _{i}^{N}\right\rangle \left\langle \psi _{i}^{N}\right| ,
\end{equation}%
where $\underset{i}{\sum }p_{i}=1$, $p_{i}>0$ and $\psi _{i}^{\alpha }$ with 
$i=0,1,\cdots $, is any normalized state of the subsystem $\alpha $. Hence,
if any multiparticle state cannot be written in the above forms, the state
is called an entangled state. However, the definition \textit{per se }is not
operational, we have to turn to an operational one.

Three-particle states can be classified according to whether they are
seperable or not with respect to the different qubits [10]. They can be
classified into five classes according to whether they can be written in one
or more the following forms [10]:

\begin{equation}
\rho =\underset{i}{\sum }\left| \psi _{i}^{1}\right\rangle \left\langle \psi
_{i}^{1}\right| \otimes \left| \psi _{i}^{2}\right\rangle \left\langle \psi
_{i}^{2}\right| \otimes \left| \psi _{i}^{3}\right\rangle \left\langle \psi
_{i}^{3}\right| ,
\end{equation}

\begin{equation}
\rho =\underset{i}{\sum }\left| \psi _{i}^{1}\right\rangle \left\langle \psi
_{i}^{1}\right| \otimes \left| \psi _{i}^{23}\right\rangle \left\langle \psi
_{i}^{23}\right| ,
\end{equation}

\begin{equation}
\rho =\underset{i}{\sum }\left| \psi _{i}^{2}\right\rangle \left\langle \psi
_{i}^{2}\right| \otimes \left| \psi _{i}^{13}\right\rangle \left\langle \psi
_{i}^{13}\right| ,
\end{equation}

\begin{equation}
\rho =\underset{i}{\sum }\left| \psi _{i}^{3}\right\rangle \left\langle \psi
_{i}^{3}\right| \otimes \left| \psi _{i}^{12}\right\rangle \left\langle \psi
_{i}^{12}\right| ,
\end{equation}%
where$\left| \psi ^{1}\right\rangle $, $\left| \psi ^{2}\right\rangle $ and $%
\left| \psi ^{3}\right\rangle $ are states of system 1,2 and 3,
respectively, and $\left| \psi ^{12}\right\rangle $,$\left| \psi
^{23}\right\rangle $ and $\left| \psi ^{13}\right\rangle $ are states of two
systems. But no matter how many classes it can be classified into, one can
describe it with three cases for convenience: 1) fully seperable states,
corresponding to (3); 2) incompletely seperable states, corresponding to
(4), (5) and (6); 3) fully inseperable states, corresponding to none of
above forms.

Considering an $N$-particle pure state no matter which is separable or not ,
one can always expand it in a series of common basis. If the dimension of
the \textit{i}th subsystem is $D_{i}$, then the dimension of the common
basis must be in $\underset{i}{\Pi }D_{i}$ dimension. Hence, an N-particle
state with the fixed dimension of every subsystem can always be converted
into a single state with much higher dimension in mathematics. Thus, we can
also express an N-particle pure state $\left| \Psi ^{ABC\cdots
N}\right\rangle $ in $s$ dimension as $\left| \Psi ^{ABC\cdots
N}\right\rangle =\underset{i}{\sum }\sqrt{\lambda _{i}}\left| \Psi
_{i}^{1}\right\rangle \otimes \left| \Psi _{i}^{2}\right\rangle $ in terms
of the generalized Schmidt decomposition [13], where $\left| \Psi
_{i}^{1}\right\rangle $ and $\left| \Psi _{i}^{2}\right\rangle $ are defined
in $n_{1}$ and $n_{2}$ dimension, respectively, with $n_{1}\times n_{2}=s$.
I.e. an N-particle pure state $\left| \Psi ^{ABC\cdots N}\right\rangle $ can
always be written as a bipartite state in form, which corresponds to the
bipartite grouping of the $N$-particle system. $\left| \Psi
_{i}^{1}\right\rangle $ and $\left| \Psi _{i}^{2}\right\rangle $ correspond
to each group, respectively. Analogously, an N-particle mixed state $\rho
^{ABC\cdots N}=\underset{i}{\sum }p_{i}\left| \psi _{i}^{A\cdots
N}\right\rangle \left\langle \psi _{i}^{A\cdots N}\right| $ can be operated
in the same way because of every pure state $\psi _{i}^{A\cdots N}$.
Therefore, considering the bipartite grouping, multiparticle(N-particle)
states can also be classified into three classes analogous to three-particle
classfication. i.e.%
\begin{equation}
\rho ^{ABC\cdots N}=\underset{i}{\sum }p_{i}\left| \psi
_{i}^{A}\right\rangle \left\langle \psi _{i}^{A}\right| \otimes \cdots
\otimes \left| \psi _{i}^{N}\right\rangle \left\langle \psi _{i}^{N}\right| ,
\end{equation}%
\begin{equation}
\rho ^{ABC\cdots N}=\underset{i}{\sum }p_{i}\left| \psi
_{i}^{\sum_{j}}\right\rangle \left\langle \psi _{i}^{\sum_{j}}\right|
\otimes \left| \psi _{i}^{\sum -\sum_{j}}\right\rangle \left\langle \psi
_{i}^{\sum -\sum_{j}}\right| ,
\end{equation}%
and%
\begin{equation}
\rho ^{ABC\cdots N}=\underset{i}{\sum }p_{i}\left| \psi _{i}^{A\cdots
N}\right\rangle \left\langle \psi _{i}^{A\cdots N}\right| ,
\end{equation}%
where $\sum_{j}=q$ denotes any $q$ subsystems among $A\cdots N$ , $\psi
_{i}^{\sum_{j}}$ stands for a common state of $\sum_{j}$ subsystems, $\left|
\psi _{i}^{\sum -\sum_{j}}\right\rangle $ stands for the common state of the
rest subsystems except the $\sum_{j}$ ones and $\left| \psi _{i}^{A\cdots
N}\right\rangle $ denotes a common fully inseparable state of all the $N$
subsystems. If we divide the N subsystems $\rho ^{ABC\cdots N}$ into two big
subsystems $\rho ^{\sum_{j}}$ and $\rho ^{\sum -\sum_{j}}$, one includes one
subsystem, i.e. $\sum_{j}=1$ denotes any one of the N subsystems, and the
other includes N-1 subsystems. It is obvious that if 
\begin{equation}
\rho ^{ABC\cdots N}=\underset{i}{\sum }p_{i}\rho _{i}^{\sum_{j}}\otimes \rho
_{i}^{\sum -\sum_{j}}
\end{equation}%
holds for every $\sum_{j}=1$ (There exist $C_{N}^{1}=N$ ways to realize such
a bipartite grouping.), then the whole N-particle system is semiseparable,
which was defined in [12]. Note, however, that if there does exist none of
all the $\sum_{j}=1$ such that\ (10) holds, we cannot draw the conclusion
that the N-particle system is fully inseparable such as a four-particle pure
state $\psi =\psi ^{+}\otimes \psi ^{+},$ where $\psi ^{+}$ is one of the
four Bell states. $\rho ^{ABC\cdots N}$ may be incompletely separable. Based
on the above study, in order to draw a conclusion to estimate whether the
N-particle system is fully inseparable, we must enhance the above condition.
I.e. $\sum_{j}$ cannot denote only one of the N subsystems, but every case
of $\sum_{j}$ $=1,2,\cdots ,\left[ \frac{N}{2}\right] $ with $\left[ \frac{N%
}{2}\right] $ =$\left\{ 
\begin{array}{cc}
N/2, & N\text{ \ is even} \\ 
(N-1)/2, & N\text{ \ is odd}%
\end{array}%
\right. $. \ However, in order to estimate the inseparability and the
semiseparability in the same criterion, we have to express the condition of
semiseparability in terms of the condition of full inseparability, which is
equivalent to the original one, whilst we express the above conditions in a
more rigorous way.

\textit{Definition1.- }The N-particle system $\rho ^{ABC\cdots N}$, which
can be divided into two big subsystems $\rho ^{\sum_{j}}$ and $\rho ^{\sum
-\sum_{j}}$ in $\underset{i=1}{\overset{\left[ \frac{N}{2}\right] }{\sum }}%
C_{N}^{i}$ ways with $C_{N}^{i}=\frac{N!}{(N-i)!i!}$ and $\sum_{j}\in
\lbrack 1,\left[ \frac{N}{2}\right] ]$, is called semiseparable, iff (10)
holds for all $\sum_{j}\in \lbrack 1,\left[ \frac{N}{2}\right] ]$, and
called fully inseparable, iff $\nexists $ $\sum_{j}\in \lbrack 1,\left[ 
\frac{N}{2}\right] ]$ such that (10) holds.

\section{Multiparticle Free Entanglement Measurement}

According to the above redefinition and analysis, we have converted the
study of multiparticle inseparabilitiy into study of a series of bipartite
inseparability in mathematics. In other words, multiparticle entanglement
measure can be obtained from a series of bipartite entanglement measures
corresponding to every bipartite grouping. However, note that it does not
mean that multiparticle entanglement is equivalent(converted to each other)
to bipartite entanglement.

Because the way to divide the whole system to two big subsystems (bipartite
grouping) is stochastic and equavilent, the entanglement measure of the
whole system is given by 
\begin{equation}
\overline{E}=\underset{j=1}{\overset{\underset{i=1}{\overset{\left[ \frac{N}{%
2}\right] }{\sum }}C_{N}^{i}}{\sum }}(E_{j}/\underset{i=1}{\overset{\left[ 
\frac{N}{2}\right] }{\sum }}C_{N}^{i}),
\end{equation}
where, $\overline{E}$ denotes the multiparticle entanglement measure of the
given system and $E_{j}$ denotes the bipartite entanglement measure
corresponding to the $j$th bipartite grouping.

For multiparticle states, one can find that all the bipartite states
obtained by our bipartite grouping are pure states or mixed states
corresponding to the original\ pure or mixed multiparticle ones. We have to
find an effective bipartite entanglement measure. For bipartite pure states,
partial entropy measure or $C(\rho )=\sqrt{2(|\left\langle \psi |\psi
\right\rangle |^{2}-Tr\rho _{r}^{2})}$ , the concurrence, which is defined
in [14], work well in arbitrary dimension. So $E_{j}=C(\rho _{j})$ or $%
E_{j}=S_{j}(\Psi ^{12})=-tr\{\rho _{1}\log \rho _{1}\}$, where $\rho
_{1}=tr_{2}\{\left| \Psi ^{12}\right\rangle \left\langle \Psi ^{12}\right|
\} $ is the reduced density matrix and the subscripts $1$ and $2$ denote the
two big subsystems $1$ and $2$ after bipartite grouping. For bipartite mixed
states in higher dimension, it is difficult to find a \ satisfactory
operational entanglement measure. Although that ''Concurrence of mixed
bipartite quantum states in arbitrary dimensions'' [11] proposed recently
sheds new light on our problem to some extent, one can also find that this
measure is complicated and inoperational. Here, for integrality, we can
temporarily employ the concurrence of bipartite mixed state in arbitrary
dimension as bipartite entanglement measure. I.e. $E_{j}=c(\rho _{_{j}})$
with $\rho _{_{j}}$ standing for the bipartite density matrix by the
bipartite grouping in the $j$th way. In some cases, in order to give an
explicit expression of entanglement measure of a state, we can also employ
the negativity $N(\rho )=\frac{||\rho ^{T_{A}}||-1}{2}$ defined in [15],
which corresponds to the absolute value of the sum of negative eigenvalues
of $\rho ^{T_{A}}$ [17]. Of course, a better bipartite entanglement measure
will be expectable and better compensative for our measure. But no matter
which measure one chooses, in the same case, one must employ the same
measure as must work without any mistake in the given case.

In our multiparticle entanglement measure, we have to divide the whole
N-particle system into two big subsystems by virtue of the above method. \
Equivalently, we can construct a series of permutation operations to realize
the bipartite grouping. Consider $\rho ^{ABC\cdots N}=\underset{i}{\sum }%
p_{i}\left| \psi _{i}^{AB\cdots N}\right\rangle \left\langle \psi
_{i}^{AB\cdots N}\right| $ in $s$ dimension, with subsystem $A$ and
subsystem $B$ in $n_{1}$ and $n_{2}$ dimension respectively, we can get that 
\begin{equation}
\rho ^{B(AC\cdots N)}=\underset{i}{\sum }p_{i}\left| \psi _{i}^{B(AC\cdots
N)}\right\rangle \left\langle \psi _{i}^{B(AC\cdots N)}\right|  \notag
\end{equation}%
\begin{equation*}
=\underset{i}{\sum }p_{i}(P(n_{1},n_{2})^{T}\otimes 1^{C\cdots N})\left|
\psi _{i}^{AB\cdots N}\right\rangle \times \left\langle \psi _{i}^{AB\cdots
N}\right| (P(n_{1},n_{2})\otimes 1^{C\cdots N})
\end{equation*}%
\begin{equation}
=(P(n_{1},n_{2})^{T}\otimes 1^{C\cdots N})\rho ^{A(BC\cdots
N)}(P(n_{1},n_{2})\otimes 1^{C\cdots N}),
\end{equation}%
where the bracket in the superscripts denotes a whole subsystem, $%
P(n_{1},n_{2})$ is permutation matrix defined as 
\begin{equation}
P(n_{1},n_{2})=\underset{i=1}{\overset{n_{1}}{\sum }}\underset{j=1}{\overset{%
n_{2}}{\sum }}E_{ij}\otimes E_{ij}^{T}=\left( 
\begin{array}{cccc}
E_{11}^{T} & E_{12}^{T} & \cdots & E_{1n_{2}}^{T} \\ 
E_{21}^{T} & E_{22}^{T} & \cdots & E_{2n_{2}}^{T} \\ 
\vdots & \vdots & \ddots & \vdots \\ 
E_{n_{1}1}^{T} & E_{n_{1}2}^{T} & \cdots & E_{n_{1}n_{2}}^{T}%
\end{array}%
\right) ,
\end{equation}%
$E_{ij}$ is a matrix in $n_{1}\times n_{2}$ dimension with subscript $ij$ \
denoting the matrix element $e_{ij\ }=1$ and the rests are zero in the
matrix $E_{ij}$. By such a transformation, an $n_{1}\times \left(
s/n_{1}\right) $ bipartite $\rho ^{A(BC\cdots N)}$ is transformed to an $%
n_{2}\times \left( s/n_{2}\right) $ bipartite $\rho ^{B(AC\cdots N)}$. In
terms of dividing the whole system in $\underset{i=1}{\overset{\left[ \frac{N%
}{2}\right] }{\sum }}C_{N}^{i}$ ways, we have to construct corresponding
permutation matrix with the same quality. Generally, we first construct a
permutation which moves the $j$th particle to the position of the $i$th one
and moves the $i$th particle to the position of the $(i+1)$th one as%
\begin{equation}
P^{\prime }(i,j)=\left( \overset{i-1}{\underset{t_{1}=1}{\otimes }}%
1^{t_{1}}\right) \otimes \left( \overset{j-1}{\underset{t_{2}=i}{\otimes }}%
p(\dim (t_{2}),\dim (t_{2}+1))\right) \otimes \left( \overset{last}{\underset%
{t_{3}=j+1}{\otimes }}1^{t_{3}}\right) ,
\end{equation}%
here $1^{\alpha }$ stands for unit matrix with the same dimension to the $%
\alpha $th particle, $\dim (i)$ denotes the dimension of the $i$th particle
and $\overset{0}{\underset{t_{1}=1}{\otimes }}1^{t_{1}}=1.$ What's more, we
require that $\overset{i+1}{\underset{t_{2}=i}{\otimes }}p(\dim (t_{2}),\dim
(t_{2}+1))=$ $p(\dim (i),\dim (i+1))\otimes p(\dim (i+1),\dim (i+2))$ and $%
\overset{i-1}{\underset{t_{1}=1}{\otimes }}1^{t_{1}}$ and $\overset{last}{%
\underset{t_{3}=j+1}{\otimes }}1^{t_{3}}$ are defined analogously.
Therefore, for the $k$th grouping, one of the two big subsystems includes $M$
\ particles each of which lies on the $X_{i}$th position, we can construct a
unitary transformation by the permutation to realize the grouping as%
\begin{equation*}
U_{k}=\overset{N}{\underset{i=1}{\Pi }}P^{\prime }(i,X_{i}),
\end{equation*}%
analogously, $\overset{j+1}{\underset{i=j}{\Pi }}P^{\prime
}(i,X_{i})=P^{\prime }(j,X_{j})\times P^{\prime }(j+1,X_{j+1})$. Note that
the order of every particle in each big subsystem does not influence the
separability \ relation between the two big subsystems. The different orders
are just like local unitary transformations. If we operate every $U_{k}$ on
the density matrix $\rho ^{ABC\cdots N}$, we get a density matrix $\rho
_{k}=U_{k}^{T}\rho ^{ABC\cdots N}U_{k}$ according to the $k$th grouping.
Thus we can get a set $\rho =\{\rho _{k}|k=0,1,\cdots \underset{i=1}{\overset%
{\left[ \frac{N}{2}\right] }{\sum }}C_{N}^{i}\}$, every element of which
corresponds to every $E_{j}$ in (10).

\section{Examples}

As examples, for pure states, consider a general three-particle pure state 
\begin{equation*}
\left| \Psi ^{ABC}\right\rangle =(C_{1}\left| 0\right\rangle
_{A}+C_{2}\left| 1\right\rangle _{A})\left| \phi _{BC}^{+}\right\rangle
+(C_{3}\left| 0\right\rangle _{A}+C_{4}\left| 1\right\rangle _{A})\left|
\phi _{BC}^{-}\right\rangle
\end{equation*}%
\begin{equation}
+(C_{5}\left| 0\right\rangle _{A}+C_{6}\left| 1\right\rangle _{A})\left|
\psi _{BC}^{+}\right\rangle +(C_{7}\left| 0\right\rangle _{A}+C_{8}\left|
1\right\rangle _{A})\left| \psi _{BC}^{-}\right\rangle ,
\end{equation}%
with $\overset{8}{\underset{i=1}{\sum }}\left| C_{i}\right| ^{2}=1$ , $%
\left| \phi _{BC}^{\pm }\right\rangle =\frac{1}{\sqrt{2}}(\left|
00\right\rangle \pm \left| 11\right\rangle )$ and $\left| \psi _{BC}^{\pm
}\right\rangle =\frac{1}{\sqrt{2}}(\left| 01\right\rangle \pm \left|
10\right\rangle )$ . By our entanglement measure, $\underset{i=1}{\overset{%
\left[ \frac{3}{2}\right] }{\sum }}C_{3}^{i}=C_{3}^{1}=3$, and 
\begin{equation*}
C(\rho _{i})=\sqrt{2(|\left\langle \psi |\psi \right\rangle |^{2}-Tr(\rho
_{i})_{r}^{2})}=\sqrt{2(1-Tr(\rho _{i})_{r}^{2})},
\end{equation*}%
with $i\ $\ denoting $A-BC$, $B-AC$ and $C-AB$ , three different groupings
respectively. Hence, we have 
\begin{equation*}
\overline{E}=\frac{1}{3}\underset{i}{\sum }C(\rho ^{i})=\frac{1}{3}[C(\rho
^{A-BC})+C(\rho ^{B-AC})+C(\rho ^{C-AB})],
\end{equation*}%
with 
\begin{equation*}
C(\rho ^{i})=\sqrt{2(1-(M_{i}^{2}+N_{i}^{2}+2P_{i}Q_{i}))},
\end{equation*}%
where%
\begin{equation*}
M_{A-BC}=\left| C_{1}\right| ^{2}+\left| C_{3}\right| ^{2}+\left|
C_{5}\right| ^{2}+\left| C_{7}\right| ^{2},
\end{equation*}%
\begin{equation*}
N_{A-BC}=\left| C_{2}\right| ^{2}+\left| C_{4}\right| ^{2}+\left|
C_{6}\right| ^{2}+\left| C_{8}\right| ^{2},
\end{equation*}%
\begin{equation*}
P_{A-BC}=C_{1}C_{2}^{\ast }+C_{3}C_{4}^{\ast }+C_{5}C_{6}^{\ast
}+C_{7}C_{8}^{\ast },
\end{equation*}%
\begin{equation*}
M_{B-AC}=\frac{1}{2}(\left| C_{1}+C_{3}\right| ^{2}+\left|
C_{2}+C_{4}\right| ^{2}+\left| C_{5}+C_{7}\right| ^{2}+\left|
C_{6}+C_{8}\right| ^{2}),
\end{equation*}%
\begin{equation*}
N_{B-AC}=\frac{1}{2}(\left| C_{1}-C_{3}\right| ^{2}+\left|
C_{2}-C_{4}\right| ^{2}+\left| C_{5}-C_{7}\right| ^{2}+\left|
C_{6}-C_{8}\right| ^{2}),
\end{equation*}%
\begin{eqnarray*}
P_{B-AC} &=&\frac{1}{2}((C_{1}+C_{3})(C_{5}-C_{7})^{\ast
}+(C_{5}+C_{7})(C_{1}-C_{3})^{\ast } \\
&&+(C_{2}+C_{4})(C_{6}-C_{8})^{\ast }+(C_{6}+C_{8})(C_{2}-C_{4})^{\ast },
\end{eqnarray*}%
\begin{equation*}
M_{C-AB}=\frac{1}{2}(\left| C_{1}+C_{3}\right| ^{2}+\left|
C_{2}+C_{4}\right| ^{2}+\left| C_{5}-C_{7}\right| ^{2}+\left|
C_{6}-C_{8}\right| ^{2}),
\end{equation*}%
\begin{equation*}
N_{C-AB}=\frac{1}{2}(\left| C_{1}-C_{3}\right| ^{2}+\left|
C_{2}-C_{4}\right| ^{2}+\left| C_{5}+C_{7}\right| ^{2}+\left|
C_{6}+C_{8}\right| ^{2}),
\end{equation*}%
\begin{eqnarray*}
P_{C-AB} &=&\frac{1}{2}((C_{1}+C_{3})(C_{5}+C_{7})^{\ast
}+(C_{5}-C_{7})(C_{1}-C_{3})^{\ast } \\
&&+(C_{2}+C_{4})(C_{6}+C_{8})^{\ast }+(C_{6}-C_{8})(C_{2}-C_{4})^{\ast },
\end{eqnarray*}%
\begin{equation*}
Q_{_{i}}=P_{i}^{\ast }.
\end{equation*}%
Therefore, we have given out the entanglement measure of all the
three-particle pure states. one can evaluate $\overline{E}$ according to the
given quantum state. For example, consider $C_{1}=C_{4}=\frac{1}{\sqrt{2}}$
and the rest are zero, then $\overline{E}=1$ by our measure, which suggests $%
\left| \Psi ^{ABC}\right\rangle $ is a GHZ state. Substitute $C_{1}$ and $%
C_{4}$ to (15), then $\left| \Psi ^{ABC}\right\rangle =\frac{1}{2}(\left|
0\right\rangle +\left| 1\right\rangle )^{A}\left| 00\right\rangle ^{BC}+%
\frac{1}{2}(\left| 0\right\rangle -\left| 1\right\rangle )^{A}\left|
11\right\rangle ^{BC}$ which is only a local unitary transformation
different from the GHZ state, $\left| \Psi ^{^{\prime }ABC}\right\rangle =%
\frac{1}{\sqrt{2}}\left| 000\right\rangle ^{ABC}+\frac{1}{\sqrt{2}}\left|
111\right\rangle ^{ABC}$. This result is consistent to our measure.

For mixed states, consider such a state as 
\begin{equation}
\rho =x\left| \psi _{0}^{+}\right\rangle \left\langle \psi _{0}^{+}\right| +%
\frac{1-x}{2^{N}}1,
\end{equation}%
which is described in [10,16]. We employ the negativity $N(\rho )$ as
entanglement measure. By our method, one can find that the $N$-particle
state can be divided in $\underset{i=1}{\overset{\left[ \frac{N}{2}\right] }{%
\sum }}C_{N}^{i}$ ways, and $N(\rho _{j})=\left| \frac{1-(1+2^{N-1})x}{2^{N}}%
\right| $ for $j=1,2,\cdots \underset{i=1}{\overset{\left[ \frac{N}{2}\right]
}{\sum }}C_{N}^{i}$ and $x>\frac{1}{(1+2^{N-1})}$ with subscript $j$ \
denoting the bipartite grouping in the $j$th way. Hence one can get 
\begin{equation}
\overline{E}(\rho )=\left\{ 
\begin{array}{cc}
\left| \frac{1-(1+2^{N-1})x}{2^{N}}\right| , & x>\frac{1}{(1+2^{N-1})} \\ 
0 & otherwise%
\end{array}%
\right. .
\end{equation}%
This result is not only consistent to the previous one [10], but also gives
an explict entanglement measure for $x>\frac{1}{(1+2^{N-1})}$.

\section{Conclusion}

In conclusion, we have shown a redefinition to estimate whether a
multiparticle system is semiseparable or fully inseparable. In terms of the
redefinition, we convert multiparticle entanglement measure into a series of
bipartite entanglement measures in mathematics and then give out the
multiparticle entanglement measure with a simple form $\overline{E}=\underset%
{j=1}{\overset{\underset{i=1}{\overset{\left[ \frac{N}{2}\right] }{\sum }}%
C_{N}^{i}}{\sum }}(E_{j}/\underset{i=1}{\overset{\left[ \frac{N}{2}\right] }{%
\sum }}C_{N}^{i})$. At last, we give two examples demonstrating that our
measure is reasonable and feasible. Due to the convenient and operational
bipartite entanglement measure for pure states, our measure for
multiparticle entanglement works better, unlike the rather cumbersome
measure for mixed states. But in some special cases, it can also work better
to employ a special measure for bipartite states, so long as the bipartite
measure can work without any mistake. What's more, we carry out all our
studies about the inseparability and semiseparability of a state in
mathematics, and we think the inseparbility measure as entanglement measure
analogous to bipartite entanglement, but it is doesn't mean that the
multiparticle entanglement and bipartite entanglement are
equivalent(converted to each other) in physics.

\section{Acknowledgement}

We thank X. X. Yi for extensive and valuable advice. This work was supported
by Ministry of Science and Technology, China, under grant No.2100CCA00700.

\end{document}